# Designing for Health Chatbots

Ahmed Fadhil[1], Gianluca Schiavo[2]

[1]University of Trento, Trento, Italy
[2]I3, Fondazione Bruno Kessler, Trento, Italy
*Corresponding authors: ahmed.fadhil@alumni.emu.edu.tr[1], gschiavo@fbk.eu[2]

## Abstract

Building conversational agents have many technical, design and linguistic challenges. Other more complex elements include using emotionally intelligent conversational agent to build trust with the individuals. In this chapter, we introduce the nature of conversational user interfaces (CUIs) for health and describe UX design principles informed by a systematic literature review of relevant research works. We analyze scientific literature in conversational interfaces and chatterbots, providing a survey of major studies and describing UX design principles and interaction patterns.

## Introduction

Conversational user interface is a software that runs simple and structurally repetitive tasks inside a messaging application (Chakrabarti 2012). Conversation is a way to share knowledge, emotion and it is part of our communication makeup. Moreover, messaging applications are becoming the new generation of digital product design after the web and mobile application. One reason behind this shift is the simplicity offered by CUIs, compared to web or mobile application (Zakos 2008). Currently, social or messaging apps are among the most popular apps in the world (Xu 2011). However, designing CUI patterns requires personalisation to fit into the application context. There are specific design requirements when working with chatbot applications. Providing a technology with services that fit with user needs should consider user demographics, chatbot domain, data interaction, dialogue structure and so forth. Recently, Artificial Intelligence (AI) has made considerable progress in various domains, especially after the rise of neural networks and deep learning, and most of this progress is devoted to industry and business cases (Trippi 1992). One important topic in this rise is designing domain specific conversational interface experiences. CUI design pattern is complicated because conversation comes with many expectations. To meet these expectations, design patterns should be personalised to be applicable into different domains. To achieve optimum customer satisfaction, it is important to study the design aspect of the product (e.g., simplicity, efficiency, decrease frictions).

The current state of conversational interface is limited in terms of established user interface design patterns. Available studies focus mainly on designing conversation agents and handling task automation (Chakrabarti 2012). Only few works have discussed design patterns for specific chatbot application domains, and they mainly remained in the scientific literature (Satu 2015, Lokman 2010, Lundqvist 2013). There has not been deep analysis and empirical discussion of specific design elements and techniques in CUI, or highlighting issues and challenges accompanying the design pattern. A standard approach is indeed required to follow when working with CUI design patterns. This includes the available design approaches and techniques and what to consider when applying them during application design and development. For example, it is unclear when bots should be text or button based, or which the best practices when designing a bot conversation are (e.g. welcome message, conversation style, message length, etc.). However, the big question remains how to use this new medium to build great user experience. Users point of view, their needs, what motivates them to seek the bot and how to create unique CUIs for users and domain, matter when designing CUIs.

In this work, we investigate the available design techniques and elements in the literature through a survey of relevant research works. We conducted a literature review of 310 articles and research papers in the context of conversational agents and extracted applied conversational UI features discussed within the literature. Two researchers reviewed the articles and checked their domain, focus, and inclusion of specific CUI features relevant to the study. The features tackled were divided into four categories, namely Bot response / conversation, User-bot interaction, Bot development and User experience. The work contributes to the CUI design patterns from UX design and existing knowledge on CUI design. This sets the roadmap to define best practices and address CUI design challenges for researchers and developers to follow when designing or developing domain specific CUIs.

## Study Design

The design principles were informed by the findings of an integrative literature review. We started the analysis in May 2017 by constructing a corpus of papers searched through Google Scholar. The initial corpus was further integrated in August 2017 with a systematic search in Scopus, IEEExplore and ACM digital libraries. Each paper, as well as each cited paper was examined to see whether it discussed any topics related to conversational agents for health, and hence added to the corpus. The resulting corpus consisted of 103 papers, as described in Table-1.

| Search engine / Digital library | Papers | Included | Excluded |
|---|---|---|---|
| Google Scholar (starting corpus) | 78 | 77 | 1 |
| Scopus | 59 | 26 | 206 |
| IEEExplore | 82 | | |
| ACM Digital Library | 91 | | |
| Total | 310 | 103 | 207 |

Table-1: Paper Analysis and Filtering

## Method

*Aims*

We investigate the scientific literature on conversational interfaces and provide a survey of major studies in the domain by highlighting their focus on UX design principles and interaction patterns for conversational interfaces in healthcare. We reviewed each research work and extracted the UX design principles discussed in terms of chatbot-user interaction, chatbot response, chatbot development, and user experience. This will help identify gaps and guide future research directions. Our goal is to support practitioners in exploring opportunities and limitations of conversational interfaces, providing empirically-supported guidelines and indications in interaction designs for conversational agents.

*Search Strategy and Criteria*

A search of chatbot related papers was conducted to construct a corpus of papers. The review started in May 2017 with a corpus of papers from Google Scholar retrieved using specific keywords such as: 'conversational user interface', 'chatbot systems', 'health bots' and 'intelligent chatbot'. The corpus was then updated in August 2017 with a systematic review from Scopus, ACM Digital Library, and IEEExplore. Papers were examined to see whether they discussed any CUIs related topics and relevant papers were added to the corpus (see Figure-4.1). The research was guided by the combination of keywords from the CUI and the UX/HCI fields. The terms used in the search was: "human-computer interaction" ‖ "computer-human interaction" ‖ "user experience" ‖ "usability" && "conversational user interface" ‖ "chatbot" ‖ "chatterbot" ‖ "agent dialogue" ‖ "health conversational agents". Only peer reviewed papers written in English were included in the analysis (empirical studies, experimental papers, experience reports). The included papers must have reported on chatbot conversational interface and user experience / human-computer interaction. All papers that did not match these criteria were excluded (for example, papers generally discussing text-based interaction, multi agent systems or robotic agents).

*Data Coding and Synthesis*

The corpus was analyzed by adopting techniques inspired by content analysis and Grounded Theory (Corbin, 2014). For each paper included in the corpus, we provided a summary of its main characteristics, contribution and identified a draft of common theoretical themes and recurrent patterns. These themes were categorized into four main groups:
- *Bot-user interaction*
- *Bot response*
- *Bot development*
- *User experience*

For each of these groups, we synthesized the collected data and identified common CUI features that were discussed in the studies (e.g., bot personality, graphical appearance, chatbot analytics, etc.). Figure-1 depicts the review process and the main features and their associated sub-features with frequency found in the corpus. We also highlight the pipeline followed to analyze the starting corpus and the rest of the corpus obtained later. We then show the process to include or exclude the selected papers into the study.

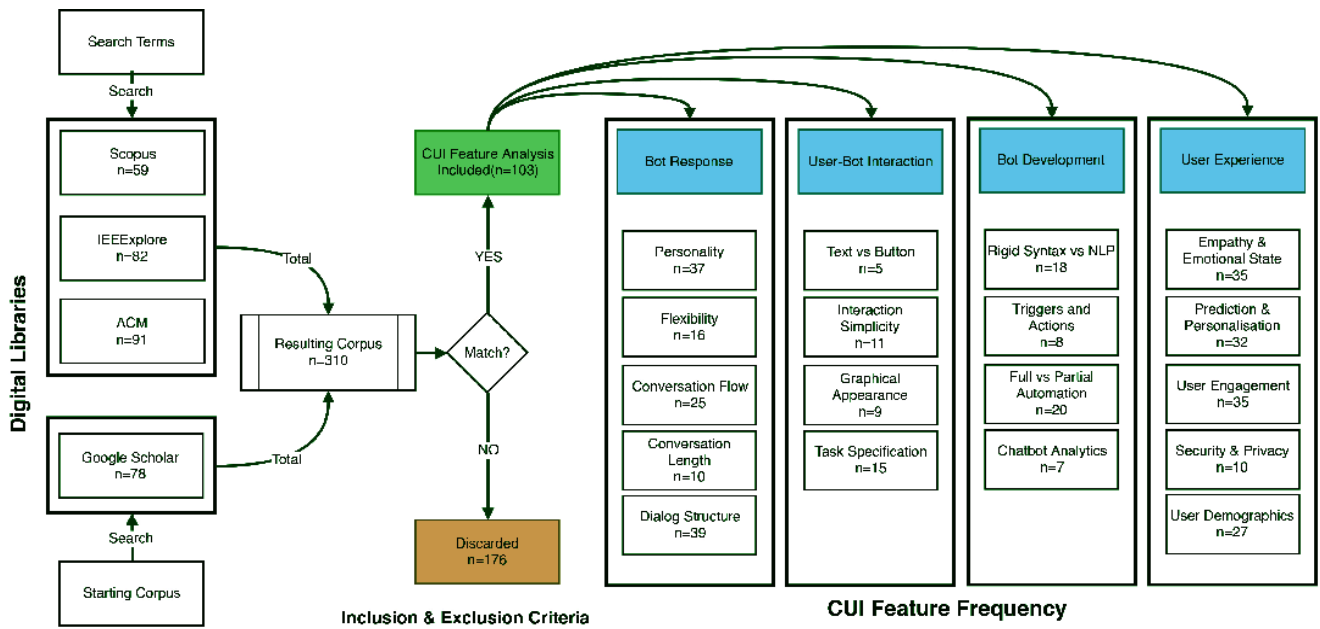

Fig-1: The Screening Process for Literature Research.

## Data Analysis
This section describes the analysis for each thematic category. Each section ends with a summary of the findings from the literature analysis.

## Bot Response
This category includes studies on user-conversation chatbots interaction, focusing on adaptive conversational behaviour expressed by conversational agents in healthcare. Chatbots are indeed gaining self-learning ability through intent classification and pattern matching to provide natural interaction. This was possible through advances in deep learning, especially in recurrent neural networks (Xing 2017). The literature analysis of user-bot interaction revealed the adoption of one or a combination of techniques in this domain (see Table-2 and Figure-2 for the feature definition and frequency).

Bot Personality: This refers to the personality that a bot assumes while interacting with the user. This feature was highlighted by several research works and covered various aspects of the user-bot interaction. The studies focused on user domination and challenge with the bot while involved in the interaction. This includes challenging the bot with irrelevant sentences (DeAngeli 2001, Brahnam 2012). User satisfaction and intention to reuse the system can help understand what engages or discourages them (Lee 2017). Users have different expectations when engaged in a conversation. Studies analysed user's personality traits inferred from their chat (Kuligowska 2015). Interactive dialogue systems were investigated in different contexts, with avatar-like chatbot (Bickmore 2005). Research based on the CASA paradigm (Nass & Moon, 2000; Reeves & Nass, 1996) has shown that people tend to respond to computers in the same manner as they would to people. Based on verbal or para-verbal cues in the bot interface, users tend to inadvertently attribute personality characteristics to the computer agents. An appropriate chatbot personality can favour engagement with the bot, such as promote feeling of natural conversation and comfort in the conversation, and even influence its persuasive intervention (Andrews, 2012). However, designing "personalities" for social agents is a complex task. Based on the review, we focused on three characteristics of bot personality:

- Character-driven dialogue. Chatbot communication is based on text. The bot must develop easy connection with users and deliver, even complex, data in a digestible way. The context should feel friendly, since users are inviting the bot into an intimate one-on-one chat space. Giving the bot an identity defines boundaries for what the bot can/cannot do. This creates content native to the bot medium, since users chat with the bot as they would with a friend. Reciprocity has been found as a strong variable in predicting relationship building between conversational agents and users (Lee 2017). Adaptation and a proactive dialogue behavior is a successful approach for supporting user-bot interaction (Langley 1999, L'Abbate et al., 2005).

- Speaking in certain voice tone. Users connect to a bot before starting a conversation: the name, profile picture and bot description contribute in creating expectations toward the conversation. Research has explored the ability of conversational agents to establish a social bond with users by displaying behaviors indicative of caring and empathy. For example, bots that use active listening techniques (Bickmore 2004, Bickmore 2005, Bickmore 2006), emphasizes commonalities with the user (Bickmore 2005), use meta-relational communication (Bickmore 2008) and refer to mutual knowledge (Bickmore 2007) are considered positive and trustworthy compared to agents that do not display such behaviors. Moreover, agents that use humor are rated as more likable, competent and cooperative than those that do not (Morkes 1998, Augello 2008).
- Speaking in one-to-one space. Research has shown that friction might arise from power differences between users and the system (the bot), especially when involved in one-to-one conversation. Users might try to exert their control over the system or show competitive attitudes toward the bot agent (DeAngeli 2002). In one-to-one interaction, the bot can tailor its interaction to the characteristics of the user. It is worth mentioning that people tend to model their communication to match that of the conversational agent (Hill 2005). The bot can adapt its language to speak in a familiar voice and increase closeness by using the name of the user or any personal information shared that can help the message. Moreover, adapting the textual interaction based on user personality traits, inferred through the chat text, can influence users' perception of the chatbot (Andrews, 2012) and their willingness to confide in and listen to the conversational agent (Ciechanowski et al., 2017, Li 2017).

  Conversational agents can solve user isolation effectively. For example, personal dialogue can provide efficient urgent intervention when a user is in need. Chatbots are used as psychiatric counseling to change aspects of user habits and lifestyle, such as drinking habit (Oh 2017, Lee et al., 2017). A study by Oh (Oh 2017) suggested a conversational service for psychiatric counseling by adapting continues emotion monitoring and methodologies to understand counseling content based on natural language understanding (NLU), and emotion recognition based on a multi-modal approach. The study highlighted chatbots inability to adapt to long-term mental diseases, since there is no continuous observation. Chatbots are also suitable health companion, they can perform diagnoses, remind medication and book doctor appointment. This is a great health monitoring tool, that is simple and always accessible (Madhu 2017). Bot personality includes knowing what the user might ask for and what the bot can't offer. There is no one-size-fits-all approach when designing correct conversation flow and currently there are no patterns that work for all cases. Most bot applications lack the structure to steer a conversation from the beginning to the end. Bot personality must define steps to guide users to learn and manage their intention. The bot should always be one step ahead of the user and infer user characteristics to adapt its response. For example, the bot should state which topics it covers when greeting the user, as below:

**Bot: I can help you track your daily diet and exercise.**

This guide results in one of the provided options being selected by the user. We need to understand what motivates users within a conversation and how this can be programmed into chatbot behavior. Bot personality must reflect the specific domain it is employing. For example, if the chatbot is collecting initial user information, then it should collect the information step-by-step, as illustrated below.

**Bot: You can insert your age below**
**User: I am 30 years old**
**Bot: Right, 30 years old, and what do you have dietary restrictions?**

Response Flexibility: Conversational agents should handle critical cases, such as outside context questions. The literature analysis found several studies that touched this aspect. They focused on human-to-human vs human-to-bot communication and the effect of both approaches (Hill 2015). User satisfaction was studied by several researchers, together with user-bot engagement (Lee 2017, Savage 2016, Jenkins et al., 2007). Interactional enjoyment and perceived trust are significant mediators in chatbot interaction (Lee 2017). To support engagement within the conversation, bots should provide flexible responses to various user requests (Jenkins et al., 2007). For example, by providing different error messages as response to the same question posed by the user. The bot should cover irregular cases, such as a keyword related to another branch of the decision tree or a keyword that is completely irrelevant to the context. If the user asks a random question or tricks the bot with unrelated questions, then it is important that the bot not repeat itself with a response such as:

**Bot: Sorry I didn't quite get that!**

If the bot continues to provide no information on alternate course of action, then the probability that a user will leave the bot is very high. A best strategy is to play with multiple types of response or error messages and see what gets the best response from the user by focusing on a humorous or satirical approach to the problem. If the error occurs when the user is asked to provide some information, such as their address, then the bot could handle that by responding:

**Bot: Sorry, I didn't get that. Please, try formatting your address: 231 Cedar St, NY,11211**

This is effective to keep user engaged and friction low. With the asynchronous chat, some replies arrive later. Having flexibility with bot interaction is essential to sustain the conversation. Therefore, bots must be present in the environment where users spend most of their time. Moreover, introducing elements, such as notification is very effective to maintain user-bot interaction. Chatbot interaction doesn't have to be limited to chat, most services provide action buttons and images which allow users to shorten their path to perform an action (e.g., purchase completion).

Conversation Flow: This refers to the user-bot back and forth conversation flow. This is the various conversation elements that should be integrated into the conversation flow. Studies increasingly investigated chatbot system accuracy and user satisfaction (Schumaker 2007). This includes the communication style for a chatbot application within a domain (Hill 2015). Some studies discussed the role of a bot to engage users in actions (Savage 2016) or use embodied conversational agents that stimulates face-to-face conversation with patients (Bickmore 2015). Automatically answering to questions and providing appropriate answers is a complex task for any NLP based system, including CUIs. Advances in machine learning and information retrieval are allowing chatbots to automatically generate responses for users' requests (Xu 2017). However, other solutions can be applied in the design of chatbot conversations to support interaction and conversation flow. The conversation path becomes important, since it can help figure out how the conversation flow should be designed (e.g., tree hierarchy). Moreover, if the user can't find an answer to his/her question, it makes sense to add an option for the user to provide feedback about the question. The point is to prevent users from getting frustrated and provide guidance instead of repeatedly saying "Sorry, I did not understand that", as below.

**User: How many activities do I have to perform today?**
**Bot: You still have the following activities to perform: act1, act2, act3, you can write the activity to report.**

Conversation Length: This is related to the amount of information delivered by bot messages and the way it is structured. In fact, the whole chatbot architecture is based on conversation flow. Even though people are communicating with the bot for longer durations, the dialogue lacks vocabulary richness (Hill 2015). Studies have used a top-down approach to provide a fixed vocabulary size and the right recommendation (Langley 1999). Dialogue research is moving towards a data-driven model approach (Griol & Molina, 2015). This might include applications, such as speech recognition, machine translation and information retrieval. Such approaches will likely come in the form of end-to-end trainable systems. This is used to convey the right information to the user and the right trigger time (Serban 2017). For example, using conversational agents for medication adherence (Bickmore 2008). Comparing how people communicate via text messages with an intelligent agent as opposed to a human, research has found that people communicate with a chatbot for longer durations, but with shorter messages (Hill 2005). User-bot interaction should be short and precise. Introducing a protracted back and forth conversation will make it feel laborious and hard to interact with the bot. Instead, bots must serve specific tasks and intend for specific domains. Unlike, GUI that defines rules for each interaction which often frustrates users, CUI must be liberating in their familiarity, e.g.,

**Bot: I can track your diet, sleep, stress and exercise. Please select your options:**
**User: Stress Management**

Dialogue structure: Dialogues should be specific to a domain and a demographic. This considers various features to structure the conversation in a way that it considers user engagement and efficiently covers the task. People tend to model their communication to match that of a chatbot (Hill 2015). Using a communication tone that fits with user preferences, emotional state, behavior and demographics increases the possibility of bot success to build the right interaction. To achieve this, researchers use several modelling languages, speech recognition and other natural language understanding tools (Ahmed 2015). The aim is to have a powerful ontology able to build communication dialogue and detect user emotional states and intents. Conversational agents power lays in them knowledge base and their ability to cover vast number of topics.

The most common chatbot knowledge representation is based on a modular knowledge representation. This makes concurrent and synergic use of different techniques, using the most adequate methodology for the management of a specific characteristic of the domain, of the dialogue, or of the user behavior (Pilato 2011). A study by Pilato et al., (Pilato 2011) built a set of modules exploiting different knowledge representation techniques and capabilities to manage conversation features. The set of modules help the chatbot to activate the most relevant knowledge section during a conversation. In a similar study by Augello et al., (Augello 2011) a web-based knowledge architecture was developed for conversational agents. This allows building specific modules to activate features of a conversation, such as topics bot covers. This approach enhances building chatbot knowledge-base that manages specific dialoguing tasks. Chatbot dialogue structure may vary depending on the context, user demographics, and the intended purpose of the bot (see also "User experience" section below). For example, a bot targeting teenagers may use less formal language and more short messages, (e.g., Bot: Hey Jami, how about we check your performance since last time?), than the one developed for the elderly (e.g., Bot: Hi Mr. Smith, how have you been since the last plan?). The dialogue should consider the data interaction the bot must acquire from the user to determine condition or provide feedback. This could include data about user behavior (Schulman 2011), personality (Li 2017), emotional and affective states (Berthelon 2013), demographic information (Bickmore 2005) or user preferences (Langley 1999). Such information can be extracted from user interaction or from existing corpora (Serban 2017).

Chatbot scripting languages, such as ChatScript and AIML (Artificial Intelligence Markup Language) are used in utterance matching against a range of dialogue patterns to produce a coherent answer following a range of responses associated to such patterns (Ahmed 2015, Mahapatra et al. 2012, Satu & Parvez, 2015). For example, the pattern could be:

**User: What was my exercise plan last month.**
**Bot: Preparing all past exercise plans...**

The system must take control to confirm a given information, classify the situation and constrain user responses. Moreover, the user can formulate a request for clarification at any time during the interaction.

| **Bot Response** ||
| --- | --- |
| **Sub-features** | **Focus** |
| *Bot Personality* | User dominance, user satisfaction, user expectation, intelligent systems to engage users |
| *Flexibility in response* | human-human vs human-bot interaction, user engagement, user satisfaction, user attitude towards agents |
| *Conversation Flow* | User engagement with bot requests, goal and non-goal driven dialogue systems, embodied conversational agents |
| *Conversation length* | Vocabulary richness, top-down approaches and bot recommendation, data-driven dialogue systems, agents for medication adherences |
| *Dialogue structure* | User communication style with the chatbot, chatbot development tools (e.g., AIML, Microsoft bot framework), Ontology and user emotion detection |

Table-2: Bot Response: Sub-features definition.

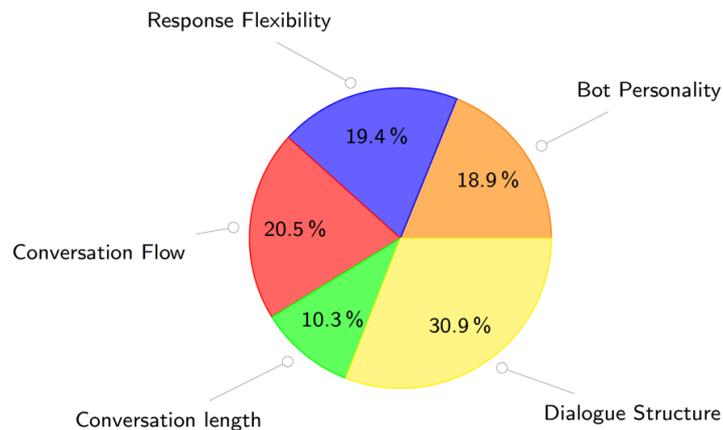

Fig. 2 Bot Response: Sub-features frequency.

**User-Bot Interaction**

Understanding user intents helps to provide relevant chatbot responses. Some approaches train the bot with actual data, however accessing relevant data is often hard. Other approaches often access the log data to analyze user intent. With the advancement in deep learning users' question can be matched with a suitable answer. However, to adapt a more AI-based approach, training the model is essential for the chatbot to connect both of those questions to correct intents and produce the correct answer. The other approach to provide bot response is by using third party services and APIs to train the chatbot (see Table-3 and Figure-3 for the feature definition and frequency).

Text vs Button: Existing chatbots are mainly text-based, however with advanced APIs and rich UI elements provided by platforms, such as Telegram, this is changing. Conversational flow uses buttons, speech-to-text or speech-to-speech, or a combination when designing bot conversations. Applying any of these depends on the context and interaction required by the user or the bot. With these techniques, bots can negotiate for an enhanced mode of interaction (Kerly 2006). User behavior is considered relevant when dealing with dialogue systems. Studies provided an overview of theories/methods in constructing or evaluating conversational agent dialogue systems (Bickmore 2006). Other studies focused on the emotional aspect of the conversation and building empathic accuracy as more important than user expressivity (Bickmore 2007). Automatic humour recognition techniques have also been applied to conversational agents to detect user humour through text or button replies (Augello 2008).

Conversational agents should overcome dead-ends where users must guess the right information. Moreover, they should support a thesaurus so to yield the same results for vocabularies, such as "buy" and "purchase". This is necessary for bots intended to negotiate with users or must respond to different vocabularies with the same intent (Kerly 2006, Klopfenstein 2017). To make user interaction easier, custom keyboards or buttons permit a limited range of inputs. However, building natural dialogue systems with rich vocabulary becomes necessary in domains, such as health, due to the sensitive data involved (Bickmore 2006, Bickmore 2007). Both buttons and texts could be applied to represent this feature. Chatbots should provide a fall-back response to guide the next steps. A proper UI should present options to the user towards a decision-making process. For instance, if the user is checking for daily meal plan, the bot can suggest a list of menus based on user preferences, as below:

**User: I want a meal plan for the dinner**
**Bot: How about menu1, menu2, menu3, click for more...**

Simplicity in interaction: This is a key when dealing with user input and providing a response to bot request. This contributes to user's long-term engagement. Simplicity is mainly focused on removing abstractions from dialogue parts and moving from general to specific conversation to detect user preferences (Langley 1999). This includes new kinds of interfaces to respond to user awareness and knowledge in chatbot application (Bickmore 2005). However, simplicity should be attentively added and consolidated argumentation which combines logic and dialectic (Grasso 2000). This includes considering user demographics, input/output approaches that fit with user ability, motivation, and action. For example, using speech dialogue as input/output can enhance user personalization and natural interaction (López-Cózar 2014).

Chatbots are effective tools in healthcare and natural language dialogues due to their simplicity in interaction (Bickmore 2005, Bickmore 2005, Grasso 2000, López-Cózar 2014). This removes the layer of abstraction introduced by interface design patterns. CUIs are great candidates for emotional considerations, perhaps even looking to add little thoughtful friction back into the interaction, so the user feels listened to.

Conversational agents should be subject specific and follow linear conversation routes, such as in education and user literacy (Bickmore 2009, Bickmore 2015, Bickmore 2009, Junior et al., 2013, Yin 2010). Users are looking for a service that finds them the answer with no extra efforts. Therefore, individual designers should avoid complicated branching paths and account for tricky failure cases. Moreover, using autocomplete, recommendations (Langley 1999), and motivational techniques, such as gamification (Vassos 2016) are effective to steer and inform the users. This shortens typing effort, since users can select an item from an autocomplete lists.

**Bot: I can track your physical activity from your phone. Click Ok if you agree.**
**User: Ok**

Graphical appearance: Just like graphical views in web or mobile applications, graphics in chatbots represent data about user performance and provide input into the bot. For example, getting a graph representation about the amount of physical activity, after acquiring data from user conversation and sensors. Graphical representation is tailored to overall design aspect of the system. This refers to pure design and application context which represents data about user health and evaluates their performance (Bickmore 2008, Bickmore 2005). However, it often refers to the way chatbots communicate with users, such as their voice tone or text typefaces (Candello 2017). This deals with chatbot and UX design through the bot development process. Graphical appearance can be the way users interact with various features (Candello 2017, Kuligowska 2015), they may include the avatar used to build natural dialogue with users, as in health dialogue systems with conversational agent for older adults (Bickmore 2006, Bickmore 2005, Bickmore 2008, Bickmore 2009). Most chatbots use either text or speech to communicate with users. Adding graphical widgets with natural language could enhance user interaction with the bot. Multi-modal communication is often preferred in complex user-system interaction mediums. Virtual agents should accurately establish the dialogue with more than a user at a time and distinguish between them. A work by Schulman et al., (Schulman 2008) described a relational agent in a science museum that is designed to conduct repeated and continuing interactions with visitors. The agent uses biometric identification system based on hand geometry, and an identification dialogue that references previous conversations. The ability to re-identify visitors enables dialogue and relationship models, with which the agent can establish social bonds with users and enhance user engagement.

Tasks Specification: Conversational agents are built for clearly defined tasks and personalized to accomplish a certain goal (Kerly 2006, Monkaresi 2013, Satu 2015). Chatbots should resemble specific character and personality and provide the right accuracy to fulfill user satisfaction (Schumaker 2007). Focusing on a topic enhances bot support for users to reflect on specific topics (Kerly 2006). Chatbots show different behavior and personality in different domains, based on their task. For example, bots in health domain usually tend to provide the right level of empathic touch with the user and use specific ontology in their conversation dialogue (Bickmore 2005, Bickmore 2011). In terms of context, bots tend to accumulate data from third party applications for producing personalized recommendations (Monkaresi 2013).
Conversational agents are personalized to accomplish a certain goal (Kerly 2006, Monkaresi 2013, Satu 2015). Since, otherwise, agents that try to do too much usually fail. For instance, Viv and Siri suffer from aiming to cover everything, and in the process compromise quality. We should understand chatbot architecture and design different bot features, considering the domain and user demographics (Neff 2016, Kuligowska 2015, Murgia 2016). Moreover, bots should convey to specific dialogue system to reflect their domain expert (Schumaker 2007, Bickmore 2005, Carroll 2000, Shawar 2007, Kusajima 2016, Klopfenstein 2017). Building an empathic relation with the user is important for long-term engagement (Bickmore 2009, Bickmore 2011, Bickmore 2010), which decides whether a user will reuse the chatbot or just abandon it all together. We illustrate below a purpose specific chatbot.

**User: Hi**
**Bot: Hi John, I can help you book/cancel a doctor's appointment and communicate with your general physician.**

| **User-Bot Interaction** ||
|---|---|
| **Sub-features** | **Focus** |
| *Text vs Button* | User negotiation, theories and methods in dialogue construction, empathic accuracy in conversation, automatic humour recognition |
| *Simplicity in interaction* | Ontology and emotional context, behaviour change intervention and user demographics |
| *Graphical appearance* | Promote health condition, user demographics, typeface perception by users |
| *Tasks & Duty specification* | System accuracy and user satisfaction, ontology and task model, app integration and user personalisation |

Table-3: User-Bot Interaction: Sub-features definition.

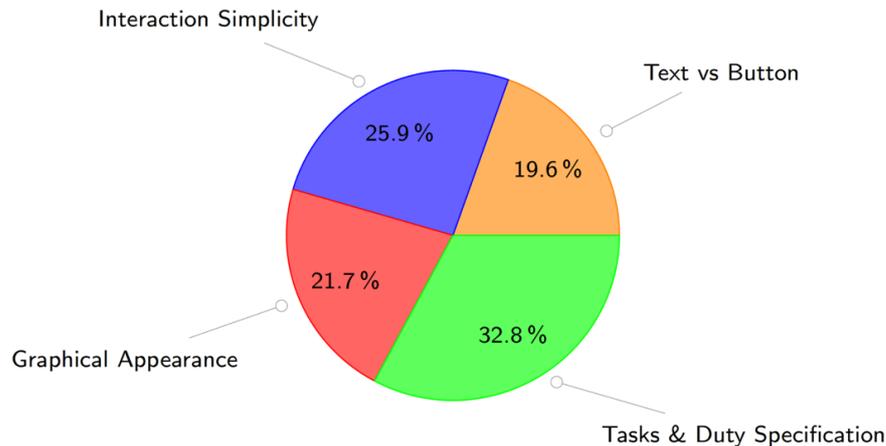

Fig. 3 User-Bot Interaction: Sub-features frequency.

## Bot Development

Feedback loops are useful to provide multiple answers to user questions. For example, to confirm user's intent, the chatbot could ask questions like "Did you mean A, B, or C". This way, users match the questions with a possible intent and the information could be used to train a machine learning model, thus improving bot accuracy. Bots relying on rule-based models can provide answers only to associated patterns. Advanced algorithms and models are required to develop advanced dialogue systems for conversational agents (see Table-4 and Figure-4 for the feature definition and frequency).

Rigid Syntax vs NLP: Chatbots function with either retrieval or generative based models. The first uses predefined rules and follows rigid syntax, whereas the second is more complicated and uses advanced deep learning models and natural language understanding (NLU). The selection of the approach depends on the type, context, and purpose of the bot and the user demographics. To provide personalized experiences, bots could integrate full conversational and knowledge-based components (Schumaker 2007). Moreover, theories and methodologies are used to construct and evaluate dialogue systems for specific purposes (Bickmore 2006, Schulman 2011). Most chatbots are retrieval-based and present some form of empathic ability. However, future dialogue research is moving towards generative models and data driven approaches (Griol and Molina, 2015). This includes speech-recognition, machine translation, and information retrieval approaches used as end-to-end trainable systems (Serban 2017). The AIML was among the most prominent chatbot languages that is based on XML. Although being effective in building retrieval based chatbots, it is limited in terms of flexibility in the dialogue. A work by Neves et al., (Neves 2003) presented XbotML language for constructing chatbot dialogues between user and chatterbot in adjacency pairs bearing one associated intention. The work contributes to the rise of fluency level of XML chatterbots by providing a linguistically grounded model for chatterbots' markup languages and allowing the extension of existing bases to different domains.

Building natural language interaction with a chatbot is still complicated. Even if NLP supports complex conversations and gets the sentence right, it will still fail in some sentences. Even if chatbots handle various user questions with smart responses, failing in some sentences might lead to user frustration. Users should not be put on hold, since this might lead to interest loss in chatting with the bot (Lookman 2010). Starting with rigid syntax, and introducing NLP later is a good approach when designing a chatbot dialogue system. Although rigid syntax has no room for error, it is relatively limited in terms of functionalities. Even though NLP is perhaps the biggest bottleneck when designing CUIs, it allows more flexibility in terms of functionalities and natural conversation. Nonetheless, users make spelling, grammar or typing mistakes. They often use slang and nuance to express ideas, which is a struggle for NLP engines. Evaluating low-level dialog systems, measuring accuracy and user satisfaction are examples of NLP complications that dialog systems encounter (Schumaker 2007, Bickmore 2006, Shawar 2002, Shawar 2007, Radziwill 2017, Marietto 2013).

Some services rely on both dialogue system and human actor to handle the conversation system. This is also due to data availability and approaches used (Serban 2017, Woudenberg 2014, Bradeško 2016, Shawar 2015, Klopfenstein 2017). Creating a good UI for dialogue systems is more important than a complex NLP system. It feels easier to click a button, as we do on apps, than to type a whole sentence that the bot might not understand. This will affect user engagement with the bot in the long-term (Bickmore 2007, Bickmore 2005). It might do the opposite of what it was asked for, to illustrate:

**User: Please stop sending me offers**
**Bot: Here are some new offers...**

Triggers and Actions: The way mobile apps and messaging platforms handle user triggers is through notification. For example, users return to their WhatsApp account after a message notification. Chatbots are no different, they provide a request to the user through notification. However, chatbots should predefine the best moment to trigger the user into a conversation. For example, calling users to achieve specific actions by providing right timing for notification (Savage 2016). Moreover, bots should examine long-term engagement techniques and user acceptance of the system. This is the case when dealing with user demographics, such as the elderly (Bickmore 2005). Dialogue complexity and using approaches in user-bot interaction are critical to be unobtrusive and create less frictions (Allen 2001).
Working on the persuasive aspect is important to trigger the user into the chatbot (e.g., motivate user to adhere to drugs). For example, when there is a steep drop-off in patient's drug adherence in the first 90 days. To persuade them to continue their medication, we should consider the associated reasons. It is hard to integrate a new routine into user's lives and a new side effect is hard to manage. Moreover, we should consider patients' needs and fears, and their demographics (Bickmore 2005), and base the design on them to fit the right goals. We need to learn what patients' value and try to connect to it with compassion and the appropriate timing. This also applies to other domains, where users are called to participate (Savage 2016, Allen 2001), illustration below.

**User: I feel depressed**
**Bot: Would you like to tell me about it, or talk to Dr. James?**

Triggering enhances reengagement with the bot. Chatbots can provide users with old conversation through a notification. However, they must set a boundary to push or pick the right time and event to trigger. Otherwise, notifications can decrease user attention span and the value intended by the service. Theory driven approaches have been shown to enhance user engagement within the conversation (Bickmore 2011, Bradeško 2016), see example below.

**Bot: Hi John, it's time you check your daily dietary plan. Open the application for details.**

Behaviour change interventions are characterised by means of a behaviour system encircled by intervention functions and by policy categories. Research is needed to establish how such approaches can lead to efficient interventions design in CUI design patterns and define the goal of various bots in different domains (Shawar 2007, Namiot 2015). We should measure how commercial applications handle re-engagement with the chatbot and the specific functionalities they embody (Kuligowska 2015).

Fully vs Partial Automation: This is tied to conversation dialogue and background services. Chatbots use three types of conversation styles, namely static, semi-automated and fully-automated conversation dialogue. The static conversation style is rule-based and does not provide any intelligence, it is simpler and easy to build. Automated refers to the generative-based model, which uses deep learning models to build interaction. This is very complex and requires a lot of training data. The semi-automated combines both or automates some parts while the rest is handled by a human. Any of the models can be used to provide enhanced negotiation and user engagement (Kerly 2006). Studies focus on conversational systems to automatically generate responses for user requests on social media by training the model to provide the right response (Xu 2017). However, in domains such as healthcare, using semi-automated approaches is more effective to handle critical cases (Woudenberg 2014). Therefore, a fully-automated system is insufficient to cover all patient needs (Bickmore 2009). Such systems should manage building natural conversation with users using topics and knowledge-bases (Woudenberg 2014). Ontologies and relational database techniques are used as approaches to drive the conversation. A work by Al-Zubaide (Al-Zubaide 2011) proposed an ontology model that maps ontologies and knowledge into a relational database. The approach overcomes the need to learn and use chatbot specific language.

Before adding functionalities, we should consider whether a chatbot is needed or a human can deliver a better service. Chatbots shouldn't replace what users are good at, rather improve what they are slow at and provide necessary negotiation facilities (Kerly 2006, Bickmore 2009, Murgia 2016). Advanced machine learning or AI will help with automation where relevant. However, a human operator is often needed in the loop to create the right end-to-end user experience. This increases quality and functionality of the chatbot (Long 2017, Kuligowska 2015, Woudenberg 2014, Shawar 2002, Shawar 2015, Lookman 2010, Bradeško 2012, Satu 2015). Chatbot developers can start with real users doing 100% of the work and later introduce role-based hierarchical trees. Afterwards, NLP should be introduced to automate 10% of common tasks. For example, NLP should be introduced for most frequently asked questions or for boarding operations. At later stages, especially as dialogue data has been acquired, more automation and intelligent machine learning techniques should be introduced (Xu 2017, Liu 2016, Namiot 2015, Jones 2016). This is effective when the value of the bot is in the human-expert, see example below.

**User: I need to book a blood test appointment**
**Bot: Hi John, I am Sam, please insert your health fiscal code to get the available dates?**

However, when the focus is task automation, starting with rigid syntax will provide helpful reaction messages to guide users when they insert incorrect formats. Starting from rigid syntax makes it easier to introduce NLP to provide flexibility in the interaction and error handling. Integrating human touch into automated replies will make users feel comfortable engaging with the chatbot or the task (Savage 2016, Bickmore 2005). A user experience is sometimes needed in the loop, especially in medical domains. This is to create the right end-to-end customer experience to provide emotional support (Bickmore 2009, Bickmore 2010), see example below.

**User: I need to book a doctor appointment**
**Bot: Hi John, please choose your date from the available dates below.**

AI Analytics: Chatbots could use other means of communication in addition to text or speech conversation. This includes representing the conversation outcome in a graph form after data analysis. This is important, since it allows to represent user performance in a single graph. Chatbots should track user data and perform analyses in the background, then represent this at the right time to the user. User-bot interaction rises different qualities of chatbot and different expectations from the users (Neff 2016). In addition, AI analytics can be applied on user emotional state. For example, chatbots can handle emotional requests from user conversation (Xu 2017). User personality traits can be inferred from chat analytics and help understand various indications about their performance (Li 2017).

Daily conversation produces a huge amount of data, which could predict future events. For example, predict what individuals are more likely to do, based on their conversation data. This is obtained after analysing user performance and digital behaviour. To ensure an accurate prediction, chatbots should perform high quality analysis of user features and performance at a certain stage (Neff 2016, Radziwill 2017). AI analytics are useful to solve or predict problems before they happen, based on data analysis. This process improves user-bot communication and could help with identifying bottlenecks, filtering conversations and understanding engagement. However, the biggest challenge in AI analytics remains the response generation, especially when it's semi-automated which introduces bias into the analysis (Xu 2017, Li 2017, Kusajima 2016). In addition, NLU and dialogue structure also play a critical role in data prediction and analysis throughout the conversation (Bickmore 2006, Bradeško 2012), see example below.

**User: Give me my medication adherence graph**
**Bot: Preparing the graph...**

| **Bot Development** ||
|---|---|
| **Sub-features** | **Focus** |
| *Rigid Syntax vs NLP* | Conversational and domain knowledge components, theories, technologies and methodologies, data-driven model, motivational interviewing technique to promote users' diet and physical activity, empathic accuracy |
| *Triggers and Actions* | Calling users for action, animated conversational agents, user demographics, spoken dialogue system |

| *Fully vs Partial Automation* | Negotiation facilities, automatic response generation, human in the loop, natural conversation with topics and knowledge-base |
| --- | --- |
| *AI Analytics* | Health and wellness data representation |

Table-4: Bot development

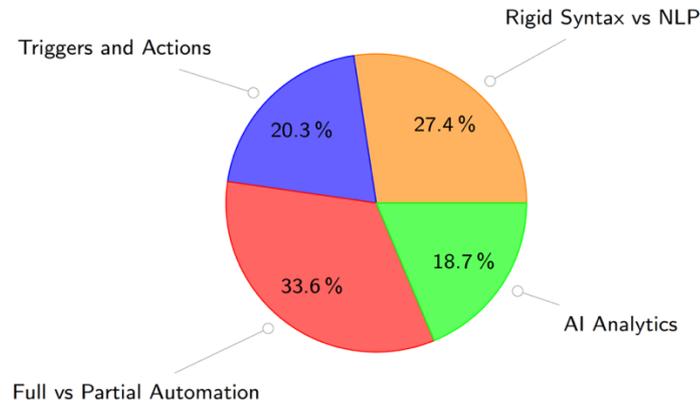

Fig. 4 Bot Development: Sub-features frequency.

## User Experience

This is mainly related to users' interaction with the bot from the design perspective. User experience is a relationship connecting user emotion and chatbot empathic ability, user demographics and personalized experiences (see Table-5 and Figure-5 for the feature definition and frequency).

Empathy and Emotional State: A strong chatbot characteristic is its ability to detect user mental state. This includes analysing user interaction through conversation logs (DeAngeli 2001) or building an ontology to detect their emotion (Berthelon 2013). To add a natural language touch, researchers have compared user-user vs user-bot interaction to examine user reaction to each technique (Hill 2015). Perceived trust and engagement are significant mediators in the relationship between variables and user satisfaction (Junior et al., 2013, Lee 2017). Lee and colleagues found that reciprocity is a strong variable compared to self-disclosure in predicting relationship building between an agent and a user (Lee 2017). A similar study focused on inferring user's psychological traits from a text-based conversation (Li 2017). One should note that user acceptance and perceived usefulness of the agent is an important aspect to predict long-term user adherence (Bickmore 2004, Bickmore 2005). Detecting mood throughout the conversation is another aspect to build empathic relation and detect user emotion (Augello 2008).

One pillar in CUI design is building empathy and approaching all design with user emotion in mind. For example, integrating pause, where users take a second to type in an answer. Chatbots perform well with decision trees and ontology to detect user intention (Berthelon 2013, Bickmore 2005). Researchers investigate adding emotional sensitivity to bots to build complicated judgments that cannot be captured without a decision tree. Addressing social issues requires emotional sensitivity and considering user behavior (Schulman 2011, Hollis 2015, Bickmore 2008, Bickmore 2010, and Yin 2010). This is a critical aspect that chatbots are universally missing. Nonetheless, even emotionless bots have changed the world for better. For instance, they have revealed epidemics of violence against young children (e.g., U-Report), or assisted healthcare providers (e.g., Ellie and Sensely) and helped users quit their bad habits (e.g., Stoptober) (Kuligowska 2015, López-Cózar 2014, Bickmore 2015, Bickmore 2007, Bennett 2010, Bickmore 2009, Bickmore 2010, Murgia 2016, Bradeško 2016, Klopfenstein 2017, Bickmore 2004). Conversational agents should pose rich dialogue vocabulary to present emotional state change (e.g., fear, anger, sadness) (Hill 2015, Long 2017, Augello 2008). For example, LawBot[1] is a legal bot created to help users understand the complexities of the law and determine whether a crime has been committed. The bot reports rape, sexual harassment, injuries and assaults, and property disputes. However, the bot relies on strict rule-based checklist to assess if a crime has been committed. If the user reports sexual harassment, and it doesn't fit within present criteria, the bot responds with the following:

**Bot: I don't think a sex offense was committed Say "Crime" for a list of what I can help you with**

---

[1] http://www.lawbot.co/

This emotional insensitivity can be counterproductive and discourage victims to speak up, thus leading to user frustration and causing friction during interaction (DeAngeli 2001, Lee 2017, Li 2017, Langley 1999, Bickmore 2005, Bickmore 2006, Pfeifer 2010, Bickmore 2009, Bickmore 2007, Bickmore 2009, Bickmore 2005, Vassos 2016). A better approach is to connect distressed users to a human expert, once the conversation has exceeded the bot's domain expertise (Walton 2014).

Prediction and Personalization: Chatbots can detect user state from their conversation and trigger the right action accordingly. Understanding user intention and feeling about a bot recommendation will help the bot predict their state and provide a personalized action. This is useful to investigate feasibility of using chatbot to support negotiation tasks combined with other systems (Kerly 2006). However, learning from user conversation is not always favourable. For example, Microsoft's chatbot Tay failed to respond to user intend, and provided nonsensical responses based on what users fed it (Neff 2016). Chatbots provide users with a linear response that are based on their states and can infer their emotional traits (Xu 2017, Li 2017). Nonetheless, bots should resemble qualities and clear functionalities that are specific for their domain (Langley 1999). For example, to explore a bot's potential to call users to action, they should be clear about what are user's role or value (Bickmore 2005, Bickmore 2006). In terms of user choice, the bot should predict what user like or dislike from their conversation and recommend an appropriate task. The main limitation of chatbots lays in their dialogue capabilities and knowledge representation. Most of the chatbots are based on simple pattern matching rules and follow a rule-based information retrieval. Analyzing distributional properties of words in a text corpus allows the creation of semantic spaces. Augello et al., (Augello 2009) presented a data-driven semantic space based on semi structured data sources freely available on the web. The chatbot exploits the semantic space to simulate an "intuitive" behaviour, attempting to retrieve semantic relations between web resources. Chatbots can ask specific and time sensitive information and obtain many variables. They can predict user preferences based on previous history and support negotiation within the interaction (Kerly 2006, Xu 2017, Li 2017, Savage 2016, Langley 1999, Bickmore 2005, Schulman 2011, López-Cózar 2014, Bickmore 2015, Bickmore 2007, Zhou 2014, abashev 2016, Kusajima 2016, Satu 2015, Vassos 2016). For instance, bots used in healthcare can ask questions that a doctor would, obtain personal information (e.g., symptoms) and provide users with accurate diagnoses. Chatbots can remember preferences, medical history and other relevant data about patients. However, users are heterogeneous with different expectations, which could affect the quality of the personal information provided (Bickmore 2006, Bickmore 2008, Bickmore 2009, Liu 2013, Bickmore 2005).

**User: I had high blood pressure today**
**Bot: Did you consider the dietary plan I gave you regarding your carbs restriction?**

User Engagement: This is an essential part of any technology. Since users are the centre of the application, without them the technology is bound to fail. Every aspect during chatbot design and development should be considered from the user's perspective. Conversational style can greatly enhance user engagement. Making users aware they are chatting with a bot is better than trying to act human (Hill 2015). User familiarity with conversational user interfaces should be considered. Some examples include interactive movie recommendations, or healthcare advice with older adults (Lee 2017, Li 2017, Bickmore 2005, Bickmore 2006). Approaches applied spoken dialogue agents to enhance user satisfaction and remove abstractions from user-bot interaction (Park 1997). Considering causes of user boredom and drop out is essential to enrich the user experience and engagement offered by the chatbot (Murgia 2016). This could be obtained through conversation or progressively simplified UI and streamlined checkout process. Studies (Candello 2017) have analyzed text typeface in user engagement and how employing different typefaces in a chatbot was perceived by users. Dialogue richness is another aspect which is often limited when dealing with user-bot conversation (Hill 2015, Murgia 2016). User satisfaction and intention to reuse virtual conversation should be evaluated to extract what engages or frustrates them (Lee 2017, Xu 2017, Long 2017, Bickmore 2005, Park 1997, Fryer 2017, morkes 1998, Bickmore 2013, Bickmore 2009). Intelligent virtual agents should offer an opportunity to enjoy the experience and transform a digital product from tolerable into friendly (Li 2017, Savage 2016, Bickmore 2006, Bickmore 2007, Bickmore 2010, Bickmore 2005). Dialogue systems for health communication rely mostly on speech dialogue style when conversing with patients (Bickmore 2005, Bickmore 2008). Behavioral theories are also used to engage users into conversing with an agent to achieve a common goal (Schulman 2011, Bickmore 2009, Bickmore 2011, Yin 2010). For example, persuasive techniques, such as gamification are used throughout the conversation to engage users with the chatbot (Vassos 2016). User engagement is also studied in the context of patient literacy and hospital discharge through a virtual agent that acts as a coach (Bickmore 2009, Bickmore 2015, Bennett 2010, Monkaresi 2013). This is important to provide timely access to health information at low cost.

While several chatbot platforms exists, there are still technical difficulties related to building data driven chatbot systems. A work by Lin et al., (Lin 2016) described an advanced platform for evaluating and annotating human-chatbot interactions. To add motivational elements to the annotation, the platform used point systems as a gamification incentive. Tasks such as chatting, evaluating, as well as annotating will grant users a certain amount of points that they can use to download previously recorded chat sessions. In the same vein, Fadhil et al., (Fadhil 2017) have discussed the approach to integrate gamification techniques, namely points and leaderboard into a chatbot application for educating teenagers about healthy diet and food waste reduction. The study discussed building conversational model and adding gamification layer on top.

Security and Privacy: Chatbots must handle sensitive information and keep it safe and secure. This is a big concern in conversational agents. For example, Tay, Microsoft's chatbot suffered from security traits that failed the bot purpose (Gulenko 2014). Security is often connected with the quality and functionality that the bot provides and the challenges it tries to overcome (Long 2017). Whereas, privacy is related to what type of user data are passed through the bot, and how it protects them. A study by Kraff and colleagues (Krafft 2017) proposed a set of guidelines for meeting the status for ethical experimentation. Privacy and security are even more critical when dealing with patients' data. A study by Bickmore et al., (Bickmore 2005) built a dialogue system for health communication with older adults and provided patients a virtual discharge nurse that interacted through face-to-face conversation. This raises many privacy issues to be handled by the agent, so that patient data will be kept confidential. Different agent qualities, technology expectations and capacities for affordance emerge in user-chatbot interaction (Neff 2016, Murgia 2016). Security and privacy are important in health domain due to the many sensitive data involved. This is the case of bots that keep users medical record about their health and future diagnoses. In the context of healthy lifestyles, the calming effects of a relational agent on users following a social bonding interaction was discussed (Bickmore 2005, Bickmore 2006, Bickmore 2009). Other measures are related to quality and functionality of conversational agents and practical and social challenges surrounding their creation. This includes the accuracy and intrusiveness of the chatbot to deliver the right information (Long 2017, Bickmore 2005, Bickmore 2013). Sharing patient data with clinicians or family raises many questions regarding privacy and social norms. Moreover, patients who suffer from chronic diseases find that self-tracking is burdensome and sharing data aggravates that burden. There are issues related to user dietary plans that are personal and chatbots should avoid being intrusive and be aware of ethical issues when dealing with sensitive user data (Krafft 2017, Bickmore 2004). This could be facilitated by allowing users to restrict information to be shared, as below:

**Bot: Hey John, would you like me to track your diet, sleep or physical activity, please confirm by selecting below**
**User: Diet & Physical activity**
**Bot: Diet, Physical activity...Great!**

User Demographic: Understanding users' background and collecting as much features about them as possible helps narrow the bot focus and provide personalized services. Before building the conversational agent, the chatbot should analyze user perception of the conversation flow and whether the users perceive it as human assistant or virtual agent (Candello 2017, Hill 2015). Chatbots should analyze what users like or dislike. This helps to avoid building a technology that cannot trigger the right action in users. For example, acceptance of animated conversational agents designed to establish long-term relationships with older adults (Bickmore 2005). User demographics are different and to build a conversational agent for them requires specific information about their treatment /disease journey. Studies (Bickmore 2010, Bickmore 2011) evaluated how hospitalized medical patients would respond to a computer animated conversational agent that would provide empathic support. Similarly, King et al., (King 2011) built an agent with a demographic target and found that a virtual advisor that delivered culturally adapted, individually tailored physical activity advice led to meaningful 4-month increase in brisk walking compared to general health education among Latino older adults. With the great personalization offered by conversational agents, it's important to cover aspects of user preferences and demographics. A study by Hijjawi (Hijjawi 2014) developed a conversational agent that supports Arabic language. The framework is based on pattern matching to handle user conversations and processes user utterance and pre-defined patterns about different topics. However, chatbots that use rule-based and general and linguistic knowledge base are limited in terms of the respond they provide. There is a need for a deeper level analysis of abstraction which can provide common sense information and result in providing more adequate responses and improve overall user experience. That said, a work by Dingli et al., (Dingli 2013) developed a framework that uses various natural language processing tools, online and offline knowledge base to enable constructing relevant responses.

Users look for people-centered approaches rather than app-centered approaches when interacting with bots. Providing alerts to users feels more like the communication we have been conditioned to have with friends or family. To provide optimum personalization and tailor the interaction to specific users, the bot should provide specific user features, such as the typeface of the system and its perception by users (Candello 2017), or the conversation flow (human-human online conversations vs human-chatbot conversations) (Hill 2015). Chatbots should examine user's intention, what motivates them, what is socially relevant and integrate them into the conversation to achieve a specific goal (Morkes 1998, Fryer 2017, Savage 2016). The acceptance and usability of an animated conversational agent designed to establish long-term relationships with older adults is an example of bots for specific demographics and age groups. The calming effect and empathic touch of a conversational agent is perceived differently by patients with different conditions (Shawar 2007, Holotescu 2016). A theory-driven computational dialogue model simulating human health counsellor is developed to help patients change via a series of conversations (Ueno 2011), and found to be associated with user demographics, such as language and conversation context. See an example of user demographics below.

**Bot: Hi John, can you type in your age please**
**User: 26**
**Bot: Cool, can you select if you have any dietary restrictions**
**User: lactose intolerance**

| **User Experience** ||
|---|---|
| **Sub-features** | **Focus** |
| *Empathy & Emotional state* | Detect user mental states, conversation logs, emotion ontology, trust and enjoyment |
| *Predict and personalization* | User state, intent, negotiation task, qualities and functionalities, user acceptance |
| *User engagement* | Typefaces, conversation style, user interest towards conversation agent, spoken dialogue |
| *Security and privacy* | Quality and functionality, user data security |
| *User demographic* | User features, users background, user preferences, user culture and ethnic, empathy with user, culturally adapted tasks |

Table-5: User Experience: Feature definition.

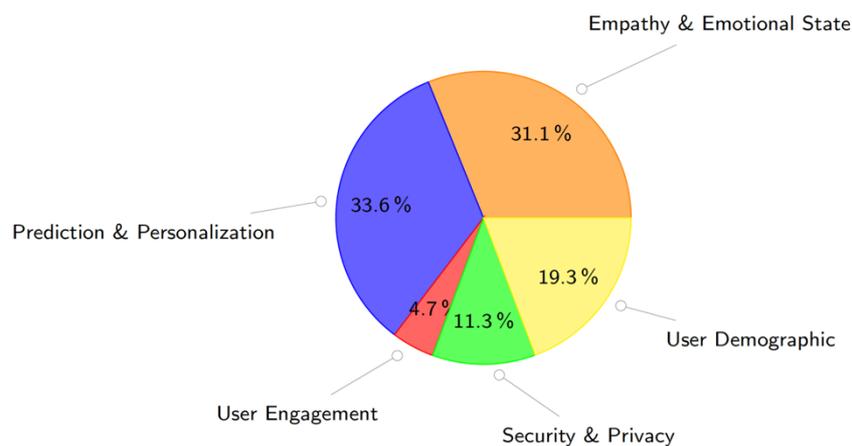

Fig. 5 User Experience: Feature frequency.

## Overall Analysis

The overall literature analysis revealed the frequency of use for some health chatbot features. Studies discussing chatbot response generation focused on dialog structure (30.9%), whereas very few discussed conversational length and dialog structure (10.3%). Studies on user bot interaction mostly discussed chatbot task specification (32.8%), and few mentioned the use of text or button in the conversation flow (19.6%). Bot development was mostly discussing chatbot dialog automation, either fully or partially (33.6%), and fewer studies in this context focused on dialog data analytics (18.7%). Finally, studies on user experience with chatbot mostly discussed user intent prediction and personalizing the conversation to fit user preferences (33.6%), fewer studies discussed user engagement with the chatbot (4.7%). In Figure-6 below we list the most and least mentioned features in these four themes of research on healthcare chatbot design.

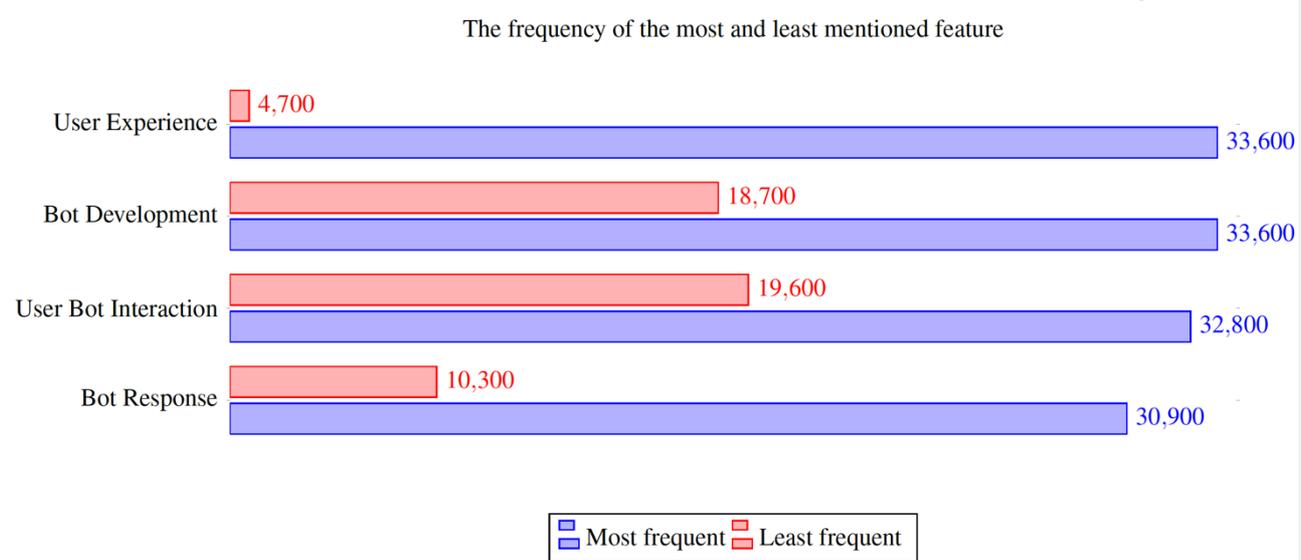

Fig. 6 Most and least feature frequency.

**Persisting Challenges**

Several limitations exist to consider within the chatbot design patterns. Not every application is suited for conversational interfaces. There are tasks that are inherently better tackled by apps, with access to local computational resources and data storage. Other tasks simply work better with a rich visual interaction that goes beyond conversational interfaces. Bots require an active internet connection to work (contrary to some modern mobile that provide offline experiences). Bots could also be a onetime use case, for example, a dermatology bot can diagnose user skin ailments and prescribe medicine or inform users about their condition. Some technical limitations of using finite state, frame-based and plan-based dialogue management techniques is that the dialog manager must manually develop the state machines with their preconditions, generic state updates and post-conditions. Dialog structures are fixed, meaning that current dialogs do not optimize future dialogs. The findings revealed no standardized way to represent all dialogs or a large range of conversational systems. Finally, few studies mentioned the role of human agent to verify the dialogue by adding contributions to the dialogue tree. This mitigates the repetitiveness in the system dialogue context which is responsible for user's engagement with the system.

**Conclusion**

In this paper, we listed common features and design elements in conversational agents for health, we then checked their coverage within the literature. Other concepts were excluded from our study, either because they fall outside the scope or require a research investigation of their own. There is a growing body of research on building and evaluating conversational agents that can negotiate and our evaluation has shown that previous works provide practical implications for the CUIs design. However, designing domain-specific CUIs represents a challenge when dealing with dialogue systems. We performed a systematic review of existing research on conversational UIs and attempted to outline important research challenges that should be addressed. Our findings revealed certain patterns that work better within a specific context. This chapter contributes to CUI research by enabling designers and developers to efficiently process, manipulate and apply best practices. Promising research directions include increasing multi-modal dialogue and properly building a sequence-to-sequence system with recurrent neural networks and considering empathic dimensions in the conversation. While much work is needed on the dialogue part of conversational agents, we hope this study will be a useful guideline to follow when designing domain specific interfaces for conversational agents.

**Reference**

[1] Chakrabarti, C., & Luger, G. F. (2012, November). A semantic architecture for artificial conversations. In Soft Computing and Intelligent Systems (SCIS) and 13th International Symposium on Advanced Intelligent Systems (ISIS), 2012 Joint 6th International Conference on (pp. 21-26). IEEE.
[2] Zakos, J., & Capper, L. (2008, October). CLIVE–an artificially intelligent chat robot for conversational language practice. In Hellenic Conference on Artificial Intelligence (pp. 437-442). Springer Berlin Heidelberg.
[3] Kenneth C Laudon, Carol Guercio Traver, and Alfonso Vidal Romero Elizondo. E-commerce. Pearson/Addison Wesley, 2007.
[4] Operator. A new & personal way to shop., 2016.
[5] LawBot. A legal bot created by cambridge university students., 2016.

[6] Nigel Walton. Chat apps will disrupt global telecoms sector. The Oxford Analytica Daily Brief, 2014.
[7] Schumaker, R. P., Ginsburg, M., Chen, H., & Liu, Y. (2007). An evaluation of the chat and knowledge delivery components of a low-level dialog system: The AZ-ALICE experiment. Decision Support Systems, 42(4), 2236-2246.
[8] De Angeli, A., Johnson, G. I., & Coventry, L. (2001, June). The unfriendly user: exploring social reactions to chatterbots. In Proceedings of The International Conference on Affective Human Factors Design, London (pp. 467-474).
[9] Candello, H., Pinhanez, C., & Figueiredo, F. (2017, May). Typefaces and the Perception of Humanness in Natural Language Chatbots. In Proceedings of the 2017 CHI Conference on Human Factors in Computing Systems (pp. 3476-3487). ACM.
[10] Hill, J., Ford, W. R., & Farreras, I. G. (2015). Real conversations with artificial intelligence: A comparison between human–human online conversations and human–chatbot conversations. Computers in Human Behavior, 49, 245-250.
[11] Kerly, A., Hall, P., & Bull, S. (2007). Bringing chatbots into education: Towards natural language negotiation of open learner models. Knowledge-Based Systems, 20(2), 177-185.
[12] Lee, S., & Choi, J. (2017). Enhancing user experience with conversational agent for movie recommendation: Effects of self-disclosure and reciprocity. International Journal of Human-Computer Studies, 103, 95-105.
[13] Neff, G., & Nagy, P. (2016). Automation, Algorithms, and Politics| Talking to Bots: Symbiotic Agency and the Case of Tay. International Journal of Communication, 10, 17.
[14] Xu, A., Liu, Z., Guo, Y., Sinha, V., & Akkiraju, R. (2017, May). A New Chatbot for Customer Service on Social Media. In Proceedings of the 2017 CHI Conference on Human Factors in Computing Systems (pp. 3506-3510). ACM.
[15] Li, J., Zhou, M. X., Yang, H., & Mark, G. (2017, March). Confiding in and Listening to Virtual Agents: The Effect of Personality. In Proceedings of the 22nd International Conference on Intelligent User Interfaces (pp. 275-286). ACM.
[16] Long, K., Vines, J., Sutton, S., Brooker, P., Feltwell, T., Kirman, B., ... & Lawson, S. (2017, May). Could You Define That in Bot Terms? Requesting, Creating and Using Bots on Reddit. In Proceedings of the 2017 CHI Conference on Human Factors in Computing Systems (pp. 3488-3500). ACM.
[17] Savage, S., Monroy-Hernandez, A., & Höllerer, T. (2016, February). Botivist: Calling Volunteers to Action using Online Bots. In Proceedings of the 19th ACM Conference on Computer-Supported Cooperative Work & Social Computing (pp. 813-822). ACM.
[18] Krafft, P. M., Macy, M., & Pentland, A. (2016). Bots as Virtual Confederates: Design and Ethics. In Proceedings of the 2017 ACM Conference on Computer Supported Cooperative Work and Social Computing (CSCW '17). ACM, New York, NY, USA, 183-190.
[19] Kuligowska, K. (2015). Commercial Chatbot: Performance Evaluation, Usability Metrics and Quality Standards of Embodied Conversational Agents. Browser Download This Paper.
[20] Ahmed, I., & Singh, S. (2015). AIML Based Voice Enabled Artificial Intelligent Chatterbot. International Journal of u-and e-Service, Science and Technology, 8(2), 375-384.
[21] Langley, P., Thompson, C., Elio, R., & Haddadi, A. (1999, July). An adaptive conversational interface for destination advice. In International Workshop on Cooperative Information Agents (pp. 347-364). Springer Berlin Heidelberg.
[22] Berthelon, F., & Sander, P. (2013, December). Emotion ontology for context awareness. In Cognitive Infocommunications (CogInfoCom), 2013 IEEE 4th International Conference on (pp. 59-64). IEEE.
[23] Bickmore, T., & Giorgino, T. (2006). Health dialog systems for patients and consumers. Journal of biomedical informatics, 39(5), 556-571.
[24] Bickmore, T. W., Caruso, L., & Clough-Gorr, K. (2005, April). Acceptance and usability of a relational agent interface by urban older adults. In CHI'05 extended abstracts on Human factors in computing systems (pp. 1212-1215). ACM.
[25] Bickmore, T. W., Caruso, L., Clough-Gorr, K., & Heeren, T. (2005). 'It's just like you talk to a friend relational agents for older adults. Interacting with Computers, 17(6), 711-735.
[26] Walker, M. A., Litman, D. J., Kamm, C. A., & Abella, A. (1997, July). PARADISE: A framework for evaluating spoken dialogue agents. In Proceedings of the eighth conference on European chapter of the Association for Computational Linguistics (pp. 271-280). Association for Computational Linguistics.
[27] Grasso, F., Cawsey, A., & Jones, R. (2000). Dialectical argumentation to solve conflicts in advice giving: a case study in the promotion of healthy nutrition. International Journal of Human-Computer Studies, 53(6), 1077-1115.
[28] Bickmore, T., & Schulman, D. (2006, April). The comforting presence of relational agents. In CHI'06 Extended Abstracts on Human Factors in Computing Systems (pp. 550-555). ACM.
[29] Serban, I. V., Lowe, R., Charlin, L., & Pineau, J. (2015). A survey of available corpora for building data-driven dialogue systems. arXiv preprint arXiv:1512.05742.
[30] Kannan, A., & Vinyals, O. (2017). Adversarial evaluation of dialogue models. arXiv preprint arXiv:1701.08198.
[31] Pfeifer, L. M., & Bickmore, T. W. (2011, March). Longitudinal Remote Follow-Up by Intelligent Conversational Agents for Post-Hospitalization Care. In AAAI Spring Symposium: AI and Health Communication.
[32] Schulman, D., Bickmore, T., & Sidner, C. (2011, March). An intelligent conversational agent for promoting long-term health behavior change using motivational interviewing. In 2011 AAAI Spring Symposium Series.
[33] Kriens, L. M. (2012). Improving medication adherence in the elderly using a medication management system (Doctoral dissertation, Tilburg University).
[34] López-Cózar, R., Callejas, Z., Griol, D., & Quesada, J. F. (2014). Review of spoken dialogue systems. Loquens, 1(2), 012.
[35] Hollis, V., Konrad, A., & Whittaker, S. (2015, April). Change of heart: emotion tracking to promote behavior change. In Proceedings of the 33rd annual ACM conference on human factors in computing systems (pp. 2643-2652). ACM.
[36] Bickmore, T., & Schulman, D. (2007, April). Practical approaches to comforting users with relational agents. In CHI'07 extended abstracts on Human factors in computing systems (pp. 2291-2296). ACM.
[37] Bickmore, T., & Pfeifer, L. (2008, April). Relational agents for antipsychotic medication adherence. In CHI'08 workshop on Technology in Mental Health.
[38] Bickmore, T., Schulman, D., & Sidner, C. (2009). Issues in designing agents for long term behavior change.
[39] Bickmore, T. W., Pfeifer, L. M., & Jack, B. W. (2009, April). Taking the time to care: empowering low health literacy hospital patients with virtual nurse agents. In Proceedings of the SIGCHI conference on human factors in computing systems (pp. 1265-1274). ACM.
[40] Bickmore, T. W., Pfeifer, L. M., Byron, D., Forsythe, S., Henault, L. E., Jack, B. W., ... & Paasche-Orlow, M. K. (2010). Usability of conversational agents by patients with inadequate health literacy: evidence from two clinical trials. Journal of health communication, 15(S2), 197-210.
[41] Woudenberg, A. V. (2014). A Chatbot Dialogue Manager-Chatbots and Dialogue Systems: A Hybrid Approach (Master's thesis, Open Universiteit Nederland).
[42] Bickmore, T., Pfeifer, L., & Paasche-Orlow, M. (2007). Health document explanation by virtual agents. In Intelligent Virtual Agents (pp. 183-196). Springer Berlin/Heidelberg.
[43] Bickmore, T., Schulman, D., & Shaw, G. (2009, September). Dtask and Litebody: Open source, standards-based tools for

| | building web-deployed embodied conversational agents. In International Workshop on Intelligent Virtual Agents (pp. 425-431). Springer Berlin Heidelberg. |
|---|---|
| [44] | Bickmore, T., Vardoulakis, L., Jack, B., & Paasche-Orlow, M. (2013, August). Automated promotion of technology acceptance by clinicians using relational agents. In International Workshop on Intelligent Virtual Agents (pp. 68-78). Springer Berlin Heidelberg. |
| [45] | Bickmore, T. W., Mitchell, S. E., Jack, B. W., Paasche-Orlow, M. K., Pfeifer, L. M., & O'Donnell, J. (2010). Response to a relational agent by hospital patients with depressive symptoms. Interacting with computers, 22(4), 289-298. |
| [46] | Bickmore, T. W., Schulman, D., & Sidner, C. L. (2011). A reusable framework for health counseling dialogue systems based on a behavioral medicine ontology. Journal of biomedical informatics, 44(2), 183-197. |
| [47] | King, A. C., Bickmore, T. W., Campero, I., Pruitt, L., & Yin, L. X. (2011). Employing'virtual advisors to promote physical activity in underserved communities: Results from the COMPASS study. Ann Behav Med, 41, S58. |
| [48] | Liu, Y., Liu, M., Wang, X., Wang, L., & Li, J. (2013, August). PAL: A Chatterbot System for Answering Domain-specific Questions. In ACL (Conference System Demonstrations) (pp. 67-72). |
| [49] | Wennberg, D. E., Marr, A., Lang, L., O'Malley, S., & Bennett, G. (2010). A randomized trial of a telephone care-management strategy. New England Journal of Medicine, 363(13), 1245-1255. |
| [50] | Bickmore, T. W., Pfeifer, L. M., & Paasche-Orlow, M. K. (2009). Using computer agents to explain medical documents to patients with low health literacy. Patient education and counseling, 75(3), 315-320. |
| [51] | Zhou, S., Bickmore, T., Paasche-Orlow, M., & Jack, B. (2014, August). Agent-user concordance and satisfaction with a virtual hospital discharge nurse. In International Conference on Intelligent Virtual Agents (pp. 528-541). Springer International Publishing. |
| [52] | Monkaresi H, Calvo RA, Pardo A, Chow K, Mullan B, Lam M, Twigg SM, Cook DI. Intelligent diabetes lifestyle coach. In OzCHI workshops programme; 2013 |
| [53] | Bickmore, T. W., Fernando, R., Ring, L., & Schulman, D. (2010). Empathic touch by relational agents. IEEE Transactions on Affective Computing, 1(1), 60-71. |
| [54] | Bickmore, T. W., & Picard, R. W. (2005). Establishing and maintaining long-term human-computer relationships. ACM Transactions on Computer-Human Interaction (TOCHI), 12(2), 293-327. |
| [55] | Yin, L., Bickmore, T., Byron, D., & Cortes, D. (2010). Cultural and linguistic adaptation of relational agents for health counseling. In Workshop on Interactive Systems in Healthcare. |
| [56] | Allen, J. F., Byron, D. K., Dzikovska, M., Ferguson, G., Galescu, L., & Stent, A. (2001). Toward conversational human-computer interaction. AI magazine, 22(4), 27. |
| [57] | Abashev, A., Grigoryev, R., Grigorian, K., & Boyko, V. (2016). Programming Tools for Messenger-Based Chatbot System Organization: Implication for Outpatient and Translational Medicines. BioNanoScience, 1-5. |
| [58] | Chu-Carroll, J., & Carberry, S. (2000). Conflict resolution in collaborative planning dialogs. International Journal of Human-Computer Studies, 53(6), 969-1015. |
| [59] | Shang, L., Lu, Z., & Li, H. (2015). Neural responding machine for short-text conversation. arXiv preprint arXiv:1503.02364. |
| [60] | Liu, C. W., Lowe, R., Serban, I. V., Noseworthy, M., Charlin, L., & Pineau, J. (2016). How NOT to evaluate your dialogue system: An empirical study of unsupervised evaluation metrics for dialogue response generation. arXiv preprint arXiv:1603.08023. |
| [61] | Shawar, B. A., & Atwell, E. (2002). A comparison between ALICE and Elizabeth chatbot systems. University of Leeds, School of Computing research report 2002.19. |
| [62] | Murgia, A., Janssens, D., Demeyer, S., & Vasilescu, B. (2016, May). Among the machines: Human-bot interaction on social Q&A websites. In Proceedings of the 2016 CHI Conference Extended Abstracts on Human Factors in Computing Systems (pp. 1272-1279). ACM. |
| [63] | Bradeško, L., Starc, J., Mladenic, D., Grobelnik, M., & Witbrock, M. (2016, September). Curious cat conversational crowd based and context aware knowledge acquisition chat bot. In Intelligent Systems (IS), 2016 IEEE 8th International Conference on (pp. 239-252). IEEE. |
| [64] | Shawar, B. A., & Atwell, E. (2007). Chatbots: are they really useful? In LDV Forum (Vol. 22, No. 1, pp. 29-49). |
| [65] | Namiot, D. (2015, November). Twitter as a transport layer platform. In Artificial Intelligence and Natural Language and Information Extraction, Social Media and Web Search FRUCT Conference (AINL-ISMW FRUCT), 2015 (pp. 46-51). IEEE. |
| [66] | Shawar, B. A., & Atwell, E. S. (2005). Using corpora in machine-learning chatbot systems. International journal of corpus linguistics, 10(4), 489-516. |
| [67] | Lokman, A. S., & Zain, J. M. (2010). Chatbot Enhanced Algorithms: A Case Study on Implementation in Bahasa Malaysia Human Language. Networked Digital Technologies, 31-44. |
| [68] | Iftene, A., & Vanderdonckt, J. (2016). MOOCBuddy: a chatbot for personalized learning with MOOCs. In Proceedings of the Rochi–International Conference on Human-Computer Interaction (p. 91). |
| [69] | Shawar, B. A., & Atwell, E. (2007, April). Different measurements metrics to evaluate a chatbot system. In Proceedings of the Workshop on Bridging the Gap: Academic and Industrial Research in Dialog Technologies (pp. 89-96). Association for Computational Linguistics. |
| [70] | Kusajima, S., & Sumi, Y. (2016, September). Twitter Bot for Activation of Online Discussion and Promotion of Understanding by Providing Related Articles. In International Conference on Collaboration Technologies (pp. 1-16). Springer Singapore. |
| [71] | Jones, S. (2015). How I Learned to Stop Worrying and Love the Bots. Social Media+ Society, 1(1), 2056305115580344. |
| [72] | Bradeško, L., & Mladenić, D. (2012). A survey of chatbot systems through a loebner prize competition. In Proceedings of Slovenian Language Technologies Society Eighth Conference of Language Technologies (pp. 34-37). |
| [73] | Klopfenstein, L. C., Delpriori, S., Malatini, S., & Bogliolo, A. (2017, June). The Rise of Bots: A Survey of Conversational Interfaces, Patterns, and Paradigms. In Proceedings of the 2017 Conference on Designing Interactive Systems (pp. 555-565). ACM. |
| [74] | Satu, M. S., & Parvez, M. H. (2015, November). Review of integrated applications with AIML based chatbot. In Computer and Information Engineering (ICCIE), 2015 1st International Conference on (pp. 87-90). IEEE. |
| [75] | Vassos, S., Malliaraki, E., Falco, F. D., Di Maggio, J., Massimetti, M., Nocentini, M. G., & Testa, A. (2016). Art-Bots: Toward Chat-Based Conversational Experiences in Museums. In Interactive Storytelling: 9th International Conference on Interactive Digital Storytelling, ICIDS 2016, Los Angeles, CA, USA, November 15–18, 2016, Proceedings 9 (pp. 433-437). Springer International Publishing. |
| [76] | Radziwill, N. M., & Benton, M. C. (2017). Evaluating Quality of Chatbots and Intelligent Conversational Agents. arXiv preprint arXiv:1704.04579. |
| [77] | Klopfenstein, L. C., & Bogliolo, A. (2017). The Quiz-Master Bot: a persistent augmented quiz delivered through online messaging. In INTED2017 Proceedings (11th International Technology, Education and Development Conference). IATED (pp. 9806-9811). |

[78] , P. (2007). Dialog management for decision processes. In Proc. of the 3rd Language and Technology Conference: Human Language Technologies as a Challenge for Computer Science and Linguistics.
[79] Marietto, M. D. G. B., de Aguiar, R. V., Barbosa, G. D. O., Botelho, W. T., Pimentel, E., França, R. D. S., & da Silva, V. L. (2013). Artificial intelligence markup language: A brief tutorial. arXiv preprint arXiv:1307.3091.
[80] Bickmore, T. W., & Picard, R. W. (2004, April). Towards caring machines. In CHI'04 extended abstracts on Human factors in computing systems (pp. 1489-1492). ACM.
[81] Morkes, J., Kernal, H. K., & Nass, C. (1998, April). Humor in task-oriented computer-mediated communication and human-computer interaction. In CHI 98 Conference Summary on Human Factors in Computing Systems (pp. 215-216). ACM.
[82] Augello, A., Saccone, G., Gaglio, S., & Pilato, G. (2008, March). Humorist bot: Bringing computational humour in a chat-bot system. In Complex, Intelligent and Software Intensive Systems, 2008. CISIS 2008. International Conference on (pp. 703-708). IEEE.
[83] Fryer, L. K., Ainley, M., Thompson, A., Gibson, A., & Sherlock, Z. (2017). Stimulating and sustaining interest in a language course: An experimental comparison of Chatbot and Human task partners. Computers in Human Behavior.
[84] Abdulkader, A., Lakshmiratan, A., & Zhang, J. (2016). Introducing deeptext: Facebook's text understanding engine.
[85] Lee, E.-J. (2008). Gender stereotyping of computers: Resource depletion or reduced attention? Journal of Communication, 58, 301–320.
[86] Sundar, S. S., & Nass, C. (2000). Source orientation in human-computer interaction: Programmer, networker, or independent social actor? Communication Research, 27, 683–703.
[87] Nass, C., & Lee, K. M. (2001). Does computer synthesized speech manifest personality? Experimental tests of recognition, similarity attraction, and consistency-attraction. Journal of Experimental Psychology: Applied, 7, 171–181.
[88] Reeves, B., & Nass, C. (1996). The media equation: How people treat computers, television, and new media like real people and places. New York: Cambridge University Press.
[89] Nass, C., & Moon, Y. (2000). Machines and mindlessness: Social responses to computers. Journal of Social Issues, 56, 81–103.
[90] Fogg, B. J., & Nass, C. I. (1997a). How users reciprocate to computers: An experiment that demonstrates behavior change. In CHI Extended Abstract (pp. 331–332). New York: ACM Press.
[91] Corbin, J., Strauss, A., & Strauss, A. L. (2014). Basics of qualitative research. Sage.
[92] Xu, Q., Erman, J., Gerber, A., Mao, Z., Pang, J., & Venkataraman, S. (2011, November). Identifying diverse usage behaviors of smartphone apps. In Proceedings of the 2011 ACM SIGCOMM conference on Internet measurement conference (pp. 329-344). ACM.
[93] Trippi, R. R., & Turban, E. (1992). Neural networks in finance and investing: Using artificial intelligence to improve real world performance. McGraw-Hill, Inc..
[94] Lundqvist, K. O., Pursey, G., & Williams, S. (2013, September). Design and implementation of conversational agents for harvesting feedback in eLearning systems. In European Conference on Technology Enhanced Learning (pp. 617-618). Springer, Berlin, Heidelberg.
[95] Katyanna Quach (September 2016). Microsoft chatbots: Sweet XiaoIce vs foul-mouthed Tay. Cultural differences, eh?. The Register. Retrieved at https://www.theregister.co.uk/2016/09/29/microsofts_chatbots_show_cultural_differences_between_the_east_and_west/
[96] Ghose, S., & Barua, J. J. (2013, May). Toward the implementation of a topic specific dialogue based natural language chatbot as an undergraduate advisor. In Informatics, Electronics & Vision (ICIEV), 2013 International Conference on(pp. 1-5). IEEE.
[97] Xing, C., Wu, W., Wu, Y., Zhou, M., Huang, Y., & Ma, W. Y. (2017). Hierarchical Recurrent Attention Network for Response Generation. arXiv preprint arXiv:1701.07149.
[98] Oh, K. J., Lee, D., Ko, B., & Choi, H. J. (2017, May). A Chatbot for Psychiatric Counselling in Mental Healthcare Service Based on Emotional Dialogue Analysis and Sentence Generation. In Mobile Data Management (MDM), 2017 18th IEEE International Conference on (pp. 371-375). IEEE.
[99] Madhu, D., Jain, C. N., Sebastain, E., Shaji, S., & Ajayakumar, A. (2017, March). A novel approach for medical assistance using trained chatbot. In Inventive Communication and Computational Technologies (ICICCT), 2017 International Conference on (pp. 243-246). IEEE.
[100] Pilato, G., Augello, A., & Gaglio, S. (2011, September). A modular architecture for adaptive chatbots. In Semantic Computing (ICSC), 2011 Fifth IEEE International Conference on (pp. 177-180). IEEE.
[101] Augello, A., Scriminaci, M., Gaglio, S., & Pilato, G. (2011, June). A modular framework for versatile conversational agent building. In Complex, Intelligent and Software Intensive Systems (CISIS), 2011 International Conference on (pp. 577-582). IEEE.
[102] Augello, A., Pilato, G., Vassallo, G., & Gaglio, S. (2009, March). A semantic layer on semi-structured data sources for intuitive chatbots. In Complex, Intelligent and Software Intensive Systems, 2009. CISIS'09. International Conference on (pp. 760-765). IEEE.
[103] Al-Zubaide, H., & Issa, A. A. (2011, November). Ontbot: Ontology based chatbot. In Innovation in Information & Communication Technology (ISIICT), 2011 Fourth International Symposium on (pp. 7-12). IEEE.
[104] Neves, A., & Barros, F. (2003). XbotML: a markup language for human computer interaction via chatterbots. Web Engineering, 329-346.
[105] Lin, L., D'Haro, L. F., & Banchs, R. (2016, October). A Web-based Platform for Collection of Human-Chatbot Interactions. In Proceedings of the Fourth International Conference on Human Agent Interaction (pp. 363-366). ACM.
[106] Fadhil, A., & Villafiorita, A. (2017, July). An Adaptive Learning with Gamification & Conversational UIs: The Rise of CiboPoliBot. In Adjunct Publication of the 25th Conference on User Modeling, Adaptation and Personalization (pp. 408-412). ACM.
[107] Hijjawi, M., Bandar, Z., Crockett, K., & Mclean, D. (2014, March). ArabChat: an Arabic Conversational Agent. In Computer Science and Information Technology (CSIT), 2014 6th International Conference on (pp. 227-237). IEEE.
[108] Dingli, A., & Scerri, D. (2013, September). Building a Hybrid: Chatterbot–Dialog System. In International Conference on Text, Speech and Dialogue (pp. 145-152). Springer, Berlin, Heidelberg.
[109] Zdravkova, K. (2000, June). Conceptual framework for an intelligent chatterbot. In Information Technology Interfaces, 2000. ITI 2000. Proceedings of the 22nd International Conference on (pp. 189-194). IEEE.
[110] Ueno, M., Mori, N., & Matsumoto, K. (2011, September). Novel chatterbot system utilizing web information for estimating current user interests. In Intelligent Data Acquisition and Advanced Computing Systems (IDAACS), 2011 IEEE 6th International Conference on (Vol. 2, pp. 656-659). IEEE.
[111] Tan, J. T. C., & Inamura, T. (2012, March). Extending chatterbot system into multimodal interaction framework with embodied contextual understanding. In Human-Robot Interaction (HRI), 2012 7th ACM/IEEE International Conference on (pp. 251-252). IEEE.

[112] Pirrone, R., Russo, G., Cannella, V., & Peri, D. (2008). GAIML: A new language for verbal and graphical interaction in chatbots. Mobile Information Systems, 4(3), 195-209.
[113] Satu, M. S., & Parvez, M. H. (2015, November). Review of integrated applications with AIML based chatbot. In Computer and Information Engineering (ICCIE), 2015 1st International Conference on (pp. 87-90). IEEE.
[114] Schulman, D., Sharma, M., & Bickmore, T. (2008, May). The identification of users by relational agents. In Proceedings of the 7th international joint conference on Autonomous agents and multiagent systems-Volume 1 (pp. 105-111). International Foundation for Autonomous Agents and Multiagent Systems.
[115] Mahapatra, R. P., Sharma, N., Trivedi, A., & Aman, C. (2012, September). Adding interactive interface to E-Government systems using AIML based chatterbots. In Software Engineering (CONSEG), 2012 CSI Sixth International Conference on (pp. 1-6). IEEE.
[116] Jenkins, M. C., Churchill, R., Cox, S., & Smith, D. (2007). Analysis of user interaction with service oriented chatbot systems. Human-Computer Interaction. HCI Intelligent Multimodal Interaction Environments, 76-83.
[117] L'Abbate, M., Thiel, U., & Kamps, T. (2005, September). Can proactive behavior turn chatterbots into conversational agents?. In Intelligent Agent Technology, IEEE/WIC/ACM International Conference on (pp. 173-179). IEEE.
[118] Boden, C., Fischer, J., Herbig, K., & Spierling, U. (2006). Citizen Talk: Application of Chatbot Infotainment to E-Democracy. Lecture notes in computer science, 4326, 370.
[119] Griol, D., & Molina, J. M. (2015). Do Human-Agent Conversations Resemble Human-Human Conversations?. In Distributed Computing and Artificial Intelligence, 12th International Conference (pp. 159-166). Springer, Cham.
[120] Junior, A. F. J., da Mata, E. C., Santana, Á. L., Francês, C. R., Costa, J. C., & Barros, F. D. A. (2013). Adapting Chatterbots' Interaction for Use in Children's Education. Emerging Research and Trends in Interactivity and the Human-Computer Interface, 413.
[121] Brahnam, S., & De Angeli, A. (2012). Gender affordances of conversational agents. Interacting with Computers, 24(3), 139-153.
[122] Andrews, P. Y. (2012). System personality and persuasion in human-computer dialogue. ACM Transactions on Interactive Intelligent Systems (TiiS), 2(2), 12.
[123] Lee, D., Oh, K. J., & Choi, H. J. (2017, February). The chatbot feels you-a counseling service using emotional response generation. In Big Data and Smart Computing (BigComp), 2017 IEEE International Conference on (pp. 437-440). IEEE.
[124] Ciechanowski, L., Przegalinska, A., & Wegner, K. (2017, July). The Necessity of New Paradigms in Measuring Human-Chatbot Interaction. In International Conference on Applied Human Factors and Ergonomics (pp. 205-214). Springer, Cham.
[125] Golder, S. A., Wilkinson, D. M., & Huberman, B. A. (2007). Rhythms of social interaction: Messaging within a massive online network. Communities and technologies 2007, 41-66.
[126] Gulenko, I. (2014). Chatbot for IT Security Training: Using Motivational Interviewing to Improve Security Behaviour. In AIST (Supplement) (pp. 7-16).